\documentclass[aps,twocolumn,amsmath,amssymb,preprintnumbers,superscriptaddress,floatfix]{revtex4}
\usepackage[utf8]{inputenc}
\usepackage{newtxtext}
\usepackage[upint]{newtxmath}
\usepackage{microtype}
\usepackage{textcomp}
\usepackage{dsfont}
\usepackage{eucal}
\usepackage{bm} 

\usepackage{enumerate}
\usepackage{amsfonts}
\usepackage{amsmath}
\usepackage{amssymb}
\usepackage{color}
\usepackage{soul}

\usepackage{graphicx}

\usepackage[colorlinks,allcolors=blue]{hyperref}
\usepackage[capitalize]{cleveref}

\renewcommand{\propto}{\sim}

\renewcommand{\d}[1]{\mathrm{d}{#1}}                            

\newcommand{\vecm}{\boldsymbol{m}}    
         
\newcommand{\vecz}{\boldsymbol{z}}

\newcommand{\etal}{\textit{et al.}}                                 

\definecolor{DarkRed}{rgb}{0.80,0,0}
\definecolor{Purple}{rgb}{0.55,0,0.55}

\renewcommand{\H}[1]{\hat{#1}}
\newcommand{\V}[1]{\check{#1}}
\newcommand{\B}[1]{\bm{#1}}

\let\phi=\varphi
\let\epsilon=\varepsilon

\newcommand{\eg}{e.g.\ }

\begin{document}
\title{Voltage control of superconducting exchange interaction and anomalous Josephson effect}
\author{Jabir Ali Ouassou}
\affiliation{Center for Quantum Spintronics, Department of Physics, Norwegian \\ University of Science and Technology, NO-7491 Trondheim, Norway}
\author{Jacob Linder}
\affiliation{Center for Quantum Spintronics, Department of Physics, Norwegian \\ University of Science and Technology, NO-7491 Trondheim, Norway}
\date{\today}
\begin{abstract}
  \noindent
  Exerting control of the magnetic exchange interaction in heterostructures is of both basic interest and has potential for use in spin-based applications relying on quantum effects.
  We here show that the sign of the exchange interaction in a spin-valve, determining whether {a parallel \textsc{(p)} or antiparallel \textsc{(ap)} magnetic} configuration is favored, can be controlled via an electric voltage.
  This occurs due to an interplay between a nonequilibrium quasiparticle distribution and the presence of spin-polarized Cooper pairs.
  Additionally, we show that a voltage-induced  distribution controls the anomalous supercurrent that occurs in magnetic Josephson junctions, obviating the challenging task to manipulate the magnetic texture of the system.
  This demonstrates that two key phenomena in superconducting spintronics, the magnetic exchange interaction and the phase shift generating the anomalous Josephson effect, can be controlled electrically.
  Our findings are of relevance for spin-based superconducting devices which in practice most likely have to be operated precisely by nonequilibrium effects.
\end{abstract}
\maketitle

\section{Introduction}
Driving a condensed matter system out of equilibrium via a control parameter such as electric voltage is a fundamentally interesting scenario.
It offers a way to alter the physical properties of the system in a controllable manner and can give rise to new types of quantum effects.
In recent years, it has been realized that rich physics ensues when considering magnetic-superconducting heterostructures that are out of equilibrium \cite{eschrig_rpp_15, beckmann_jpcm_16, bergeret_arxiv_17}.
This includes very large thermoelectric effects~\cite{kalenkov_prl_12, machon_prl_13, kolenda_prl_16}, large quasiparticle spin Hall effects~\cite{wakamura_natmat_15}, raising the paramagnetic limit of superconducting films \cite{ouassou_arxiv_18, bobkova_prb_11}, and supercurrent-induced magnetization dynamics~\cite{waintal_prb_02, zhao_prb_08, linder_prb_11, holmqvist_prb_12, takashima_prb_17, bobkova_prb_18}.
The study of such effects is associated with the field of superconducting spintronics~\cite{linder_nphys_15}, where the aim is to create a synergy between spin-polarized order and superconductivity.

Historically, creating a nonequilibrium distribution of quasiparticle states in superconducting structures has been shown to give rise to interesting effects.
A prominent example is the supercurrent transistor demonstrated in~Ref.~\cite{baselmans_nature_99}, where the direction of a Josephson effect (charge supercurrent) was tuned via a voltage-induced nonequilibrium distribution in a superconductor/normal-metal/superconductor junction~\cite{wilhelm_prl_98,Volkov1995}.
In this Letter, we explore a spin-analogue of this effect.
More precisely, we pose the question: can a \textit{spin supercurrent} be controlled via the nonequilibrium mode induced by an electric voltage? Such a spin supercurrent exists when magnetic layers are added to the Josephson junction above and physically represents the exchange interaction between these layers~\cite{slon_prb_89, nogueira_epl_04}.
If the spin supercurrent---and in particular its sign---is controlled by a nonequilibrium distribution function, it allows the preferred magnetic configuration to be switched by an electric voltage.
We show that this is indeed possible, and that it only requires {small voltages} below the superconducting gap~$\Delta$.

Additionally, we show that the recently experimentally observed anomalous phase shift in Josephson {junctions~\cite{szombati_nphys_15}} can be tuned via a nonequilibrium distribution of quasiparticles.
This is induced via an electric current and permits a nonmagnetic way to control the anomalous Josephson effect, which removes the challenging requirement to manipulate the intricate {noncollinear} magnetic texture of structures that exhibit an anomalous supercurrent~\cite{glick_sciadv_18}.
We predict large phase shifts that can be tuned by more than $\pi/2$ for voltages smaller than the superconducting gap ($\sim\!1$~meV). 
{This is two orders of magnitude smaller than the electric gate voltage that was used in Ref.~\cite{szombati_nphys_15} to observe the anomalous phase shift.}

\section{Methodology} 
\begin{figure}[t!]
  \centering
  \includegraphics[width=0.95\columnwidth]{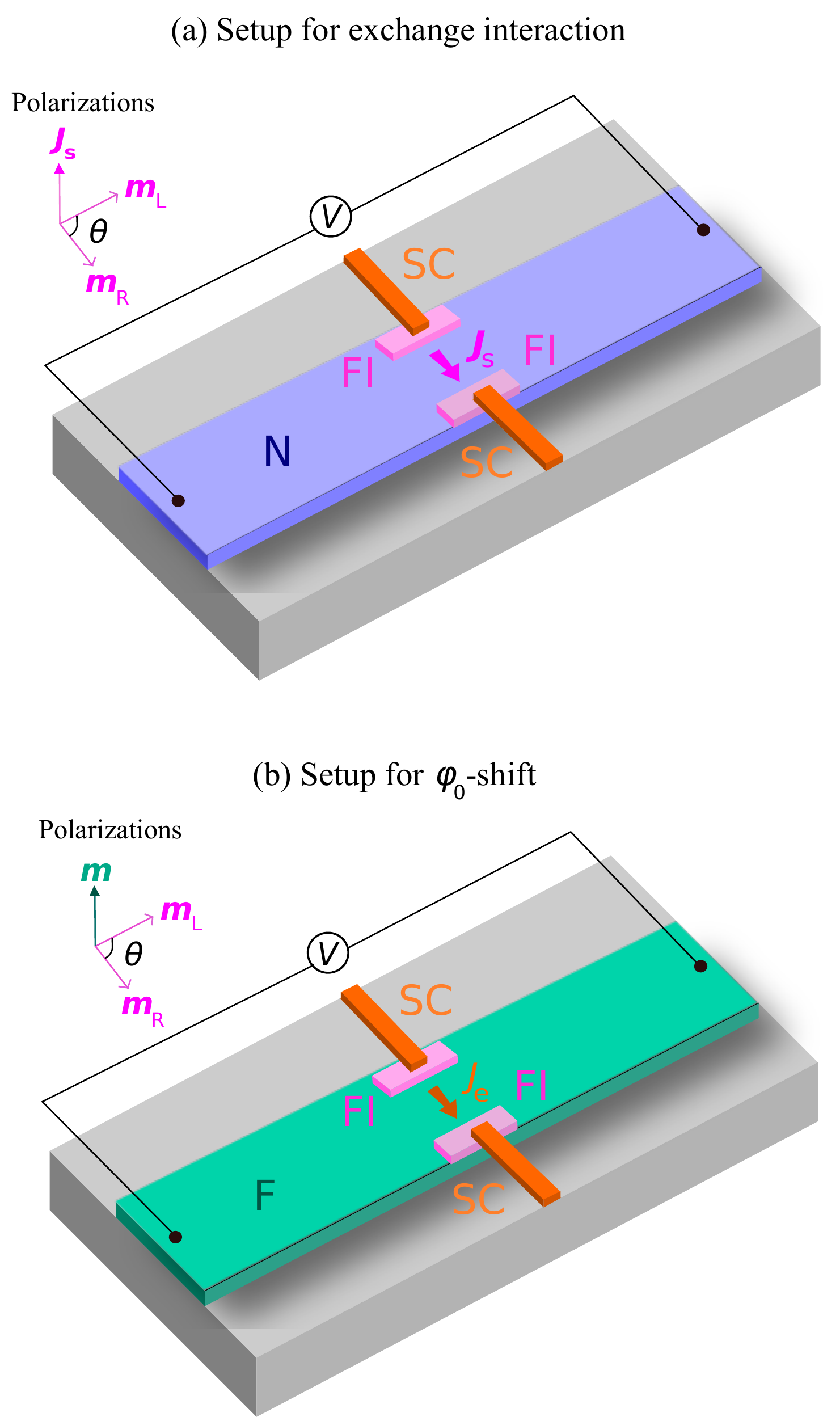}
  \caption%
  {%
      (Color online)
      Suggested experimental setups.
      External voltage sources inject resistive charge currents into normal-metal regions.
      Near the centers of these wires, there is no net charge accumulation, but an excess of \emph{both} electrons and holes compared to the equilibrium situation.
      These regions of the normal metals are then used as the weak links of magnetic Josephson junctions.
      (a)~If the weak link is a normal metal, a spontaneous spin supercurrent~$\B{J}_\text{s} \sim \B{m}_\textsc{l}\times\B{m}_\textsc{r}$ flows between the magnetic insulators (black arrow), where $\B{m}_\textsc{l}$ and $\B{m}_\textsc{r}$ refer to their magnetization directions.
      We show that this spin supercurrent can be reversed as a function of voltage, resulting in a voltage-controllable switching from {anti- to ferro}magnetic interactions between the magnets.
      (b)~If the weak link is a ferromagnet, there will in addition be a spontaneous charge supercurrent~$J_\text{e} \sim \B{m}\cdot(\B{m}_\textsc{l}\times\B{m}_\textsc{r})$ flowing between the superconductors (black arrow), where $\B{m}$ is the exchange field in the weak link.
      We show that this charge supercurrent can be tuned as a function of voltage, resulting in a voltage-controllable ground-state phase shift between the two superconductors.
  }
  \label{fig:model} 
\end{figure}

To determine the influence of nonequilibrium quasiparticle occupation in the system induced by an electric voltage, we use the quasiclassical theory of superconductivity.
This framework is well-suited to address a range of physical phenomena occuring in mesoscopic heterostructures, including charge  and spin supercurrents.
We propose experimental setups for observing our predictions in \cref{fig:model}.
These setups should be experimentally feasible as they are similar to the setup used by Baselmans \etal~\cite{baselmans_nature_99}, but with the addition of magnetic layers.
In \cref{fig:model}(a), an electric voltage injects a resistive charge current into a normal-metal wire.
At the center of each wire, there is no net charge accumulation, but a surplus of \emph{both} electrons and holes compared to the equilibrium situation.
The superconducting and normal regions are interfaced by magnetic insulators, which influence each other via an exchange interaction.
The quasiparticle injection described above alters the occupation of not only charge{-}supercurrent-carrying states, as discussed in Refs.~\cite{wilhelm_prl_98, baselmans_nature_99, heikkila_epl_00}, but also spin{-}supercurrent-carrying states, which determine the exchange interaction between the {magnets}.

In \cref{fig:model}(b), the weak link is made from a ferromagnetic metal, but except for that, the setup is identical.
When the magnetizations of the ferromagnetic insulators, $\vecm_\textsc{l}$ and $\vecm_\textsc{r}$, form a nonzero spin chirality $\chi$ together with the magnetization of the metallic ferromagnet according to $\chi = {\vecm \cdot (\vecm_\textsc{l} \times \vecm_\textsc{r})}$, an anomalous Josephson effect appears at zero phase difference between the superconductors.
This phenomenon can be understood from the fact that the broken spin-degeneracy combined with the broken chirality symmetry of the system allows the Cooper pairs to gain a net additional phase $\phi_0$ as they tunnel through the system.
By using quasiparticle injection to change the occupation of charge{-}supercurrent-carrying states, {we show below that} this anomalous Josephson current can be altered.

Both systems in \cref{fig:model} can be described by the Usadel equation for diffusive systems~\cite{bergeret_arxiv_17,Chandrasekhar2008,Belzig1999,Rammer1986,Usadel1970},
\begin{align}
  {\nabla\cdot\V{\B{I}}} &= {i}[\H{\Delta} + \B{m}\cdot\H{\B{\sigma}} + \epsilon\H{\tau}_3 ,\, \V{g}], &
  {\V{\B{I}}}            &= { -D\V{g}\nabla\V{g}, }
\end{align}
which determines the $8\times8$ quasiclassical Green function{s}
\begin{align}
  \V{g} &=
  \begin{pmatrix}
    \H{g}^\textsc{r} & \H{g}^\textsc{k} \\
    0   & \H{g}^\textsc{a}
  \end{pmatrix}, &
  \V{I} &=
  \begin{pmatrix}
    \H{I}^\textsc{r} & \H{I}^\textsc{k} \\
    0   & \H{I}^\textsc{a}
  \end{pmatrix}.
\end{align}
Above, $\H{\Delta} = \text{antidiag}(+\Delta,-\Delta,+\Delta^*,-\Delta^*)$, {$\H{\B{\sigma}} = (\H{\sigma}_1, \H{\sigma}_2, \H{\sigma}_3)$, $\H{\sigma}_n = \text{diag}(\sigma_n, \sigma_n^*)$,} $\sigma_n$ are Pauli matrices in spin space, and $\H{\tau}_n$ are Pauli matrices in Nambu space.
The parameter~$\Delta$ is the superconducting gap, which we take to be $\Delta_0 e^{\pm i\phi/2}$ for the superconductors in \cref{fig:model}, where $\Delta_0$ is the zero-temperature bulk gap, and $\phi$ is the phase difference between them.
The parameter~$\B{m}$ is the exchange field of a magnetic metal, which we take to be homogeneous. 
{We consider weakly polarized ferromagnetic alloys such as PdNi with a low content of Ni, where the exchange field is of order $10$~meV~\cite{kontos_prl_01}.}
Finally, {$D$~is the diffusion coefficient and $\epsilon$ the quasiparticle energy. 
We also define the coherence length $\xi = \sqrt{D/\Delta_0}$ and material length~$L$.}

The components of the Usadel equation are related by the identities $\H{g}^\textsc{k} = \H{g}^\textsc{r} \H{h} - \H{h} \H{g}^\textsc{a}$ and $\H{g}^\textsc{a} = -\H{\tau}_3 \H{g}^{\textsc{r}\dagger} \H{\tau}_3$, which means that it is in general sufficient to solve for the retarded component~$\H{g}^\textsc{r}$ and a distribution function $\H{h}$.
We numerically solved the equations for the retarded component using a Riccati-parametrization~\cite{Schopohl,SpinOrbit}.
The magnetic insulators in \cref{fig:model}(a--b) were treated as {spin-active tunneling boundary conditions to superconducting reservoirs}~\cite{Halfmetal, Eschrig2015b, machon_prl_13, Cottet2009, Cottet2007a},
{
  \begin{equation}
    (2L/D)\, \V{\B{I}}\cdot\B{n} = (G_\textsc{t}/G_\textsc{n}) \, [\V{g}, F(\V{g}_\textsc{bcs})] - i(G_\phi/G_\textsc{n}) \, [\V{g}, \H{m}],
  \end{equation}
  where the spin-filtering function~\cite{Halfmetal}
  \begin{equation}
    F(\V{v}) = \V{v} + \frac{P}{1+\sqrt{1-P^2}} \{ \V{v}, \H{m} \} + \frac{1 - \sqrt{1-P^2}}{1 + \sqrt{1-P^2}} \H{m} \V{v} \H{m}.
  \end{equation}
  {In the equations above, there are some products between matrices of seemingly incompatible dimensions.
  For instance, $\V{\nu}$ is an $8\times8$ matrix in Keldysh$\otimes$Nambu$\otimes$spin space, while $\H{m}$ is just a $4\times4$ matrix in Nambu$\otimes$spin space.
  Such conflicts should implicitly be resolved by taking Kronecker products with appropriate identity matrices; in this case, $\V{\nu} \H{m}$ should be interpreted as $\V{\nu} (\rho_0 \otimes \H{m})$, where $\rho_0 = \text{diag}(1,1)$ is an identity matrix in the Keldysh subspace.
  In these equations,} $\V{g}_\textsc{bcs}$ is the standard solution for a bulk superconductor~\cite{SpinOrbit}, since we treat the superconductors as reservoirs.
  Numerical values for the Drude conductance~$G_\textsc{n}$, tunneling conductance~$G_\textsc{t}$, spin-mixing conductance~$G_\varphi$, and polarization~$P$ are given in the captions of \cref{fig:exchange,fig:phi0}.
  Finally, $\B{n}$ is the interface normal, and $\H{m} = \H{\B{\sigma}}\cdot\B{m}_\textsc{l,r}$ is related to the interface magnetization.
  We use the notations $\B{m}_\textsc{l}$ and $\B{m}_\textsc{r}$ for the magnetizations of the ``left'' and ``right'' interfaces, respectively.
}
{The misalignment $\theta$  between the directions of $\B{m}_\textsc{l}$ and $\B{m}_\textsc{r}$ controls the magnitude {of the spin supercurrent} according to $\B{J}_\text{s} \propto \B{m}_\textsc{l} \times \B{m}_\textsc{r} \sim \sin\theta$. In our calculations, we have set $\B{m}_\textsc{l} = \H{\B{x}}$ and $\B{m}_\textsc{r} = \H{\B{y}}$ so that {the polarization} $\B{J}_s \propto \H{\B{z}}$.
As for the distribution function~$\H{h}$, we did not need to explicitly solve the kinetic equations~\cite{ouassou_arxiv_18, Bobkova2015a, Silaev2015, bergeret_arxiv_17, Aikebaier, Chandrasekhar2008, Belzig1999}, since an analytical solution is already known~\cite{wilhelm_prl_98, heikkila_epl_00, Pothier1997}:
\begin{equation}
  \H{h} = \dfrac12 { \{ [\tanh[(\epsilon + eV/2)/2T] \\ + \tanh[(\epsilon-eV/2)/2T] \} } \H{\tau}_0 \H{\sigma}_0. 
  \label{eq:dist}
\end{equation}
This result is valid near the centers of voltage-biased normal metals, including the weak links shown in \cref{fig:model}.
{All calculations shown herein were performed at temperature~$T=0$.
Qualitatively, we have numerically confirmed that both effects persist at all temperatures up to the critical temperature~$T_c$.
Quantitatively, the spin currents in \cref{fig:exchange} decay exponentially with temperature, being reduced by roughly one order of magnitude at $T=T_\text{c}/2$.
Moreover, the voltage required to switch the spin current direction is roughly twice as large at $T=T_\text{c}/2$.
Similarly, the $\varphi_0$ shift in \cref{fig:phi0} starts to decay quite rapidly above $T=T_\text{c}/2$.
For both setups, it is therefore beneficial to perform the experiments at low temperatures.}

The charge and spin currents were determined from the numerically calculated Green functions using standard formulas~\cite{Spinject,bergeret_arxiv_17,eschrig_rpp_15,Rammer1986,Belzig1999,Chandrasekhar2008}.
{
More precisely, distribution function in the Josephson weak link determines the sign of the superconducting spin current that mediates the exchange interaction.
The plot shows the spin supercurrent polarized in the $\vecm_\textsc{l} \times \vecm_\textsc{r} = \vecz$ direction.
The spin supercurrent drops approximately linearly from its maximum at zero to its minimum occuring at $eV/2 \approx 0.35\Delta_0$ and \textit{changes sign} in between.
As a result, the favored configuration of the magnetic insulators is changed from {anti- to ferro}magnetic by modifying the distribution of quasiparticles in the weak link with a voltage {that is smaller than} the superconducting gap. This corresponds to a voltage less than $\sim\! 1$~meV. 
We expect that the same sign reversal of the spin supercurrent should be possible when using thin ferromagnetic metals rather than ferromagnetic insulators. The stability of a given magnetic configuration at a fixed voltage is determined by the sign of the spin supercurrent, because the sign determines the direction of the torque acting on the magnetic order parameter in the ferromagnetic insulators. If the torque favors an {\textsc{ap}} configuration for one particular sign, it favors the {\textsc{p}} configuration for the opposite sign.

Moreover, we have numerically confirmed that the sign change of the spin supercurrent as a function of applied voltage occurs for a wide parameter range{, as shown in \cref{fig:exchange}(c--f).
In general, a high tunneling conductance~$G_{\textsc{t}}$ and short junction length~$L$ enhances the proximity effect; this increases the spin supercurrent at all voltages, but also increases the switching voltage required.
The polarization~$P$ has a relatively small effect on our results.
However, in the limit~$P \rightarrow 1$, it suppresses tunneling of opposite-spin Cooper pairs from the superconducting reservoirs, which is detrimental to the spin supercurrent.
Interestingly, the voltages where the strongest ferromagnetic and antiferromagnetic interactions occur are found to increase nearly linearly with the spin-mixing conductance~$G_\varphi$.
We note that for most parameter combinations explored here, a switching between \textsc{p} and \textsc{ap} ground-state configurations can be achieved using reasonable applied voltages $eV \lesssim \Delta_0$.}
{Note also that the results in \cref{fig:exchange} were calculated for zero phase difference between the superconductors, which for the setup in \cref{fig:model}(a) results in zero charge supercurrent.
This implies that the sign change in the spin supercurrent is not trivially related to any voltage-induced sign changes in the charge supercurrent.}

\begin{figure}[t!]
\centering
\includegraphics[width=\columnwidth]{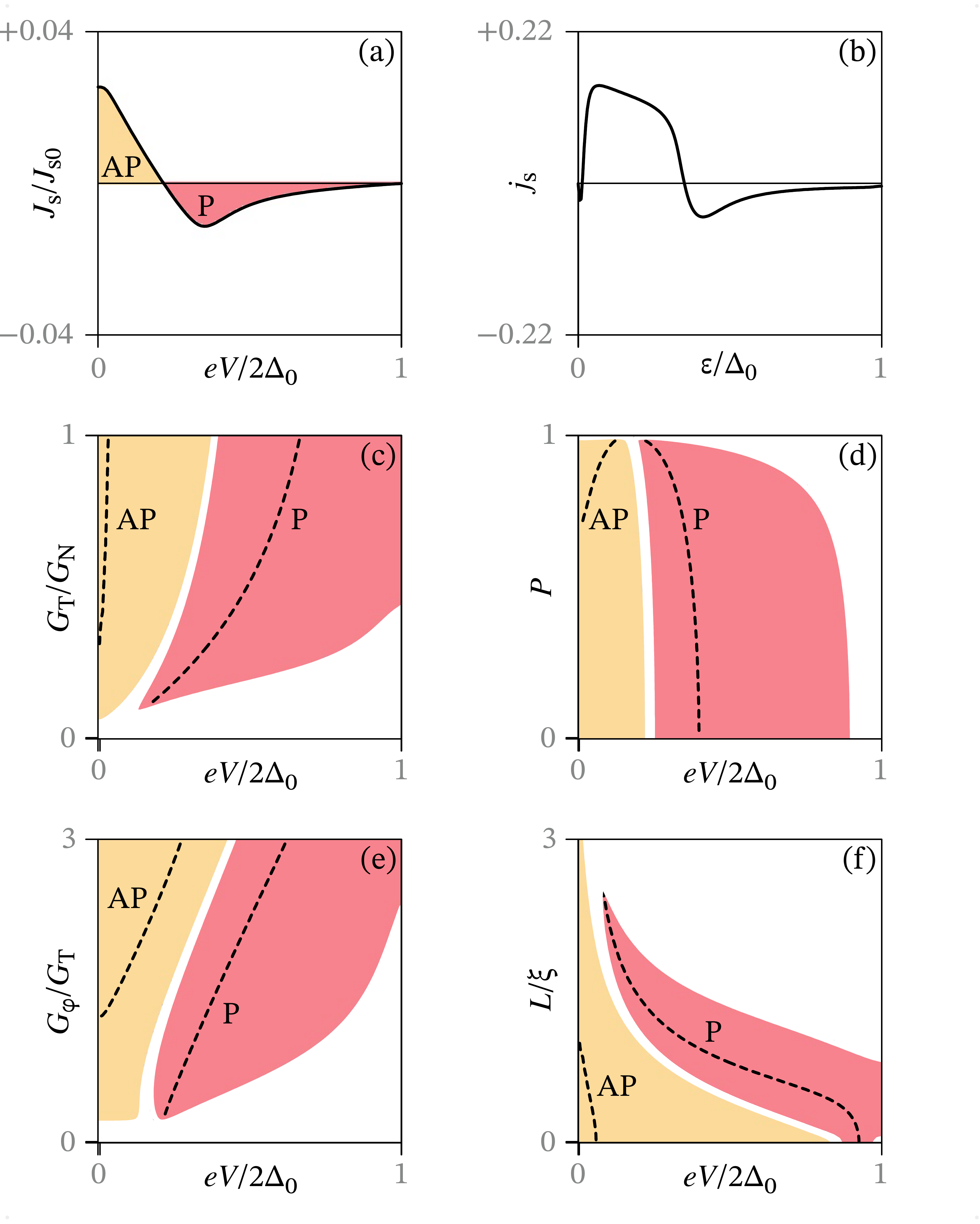}
\caption%
{%
  (Color online) Numerical results for the setup in \cref{fig:model}(a).
  (a)~Spin supercurrent~$\B{J}_\text{s}$ as a function of applied voltage.
  The voltage shifts the distribution function in the weak link of the Josephson junction, and causes the spin supercurrent to change sign at~$eV/2$ well below~$\Delta_0$.
  Since the spin supercurrent mediates the exchange interaction between the magnets, the sign reversal implies a switch from {anti- to ferro}magnetic interactions{, thus switching from a{n antiparallel~\textsc{(ap)} to parallel~\textsc{(p)}} configuration as the magnetic ground state}.
  The magnetic insulators were modeled as spin-active interfaces with polarization $P = 70\%$, tunneling conductance~$G_\textsc{t} = 0.3G_\textsc{n}$, and spin-mixing conductance~$G_\phi = 1.25 G_\textsc{t}$.
  The normal-metal weak link has a length $L_\textsc{n}=\xi$ and conductance $G_\textsc{n}$.
  (b)~Spectral spin supercurrent~$\B{j}_\text{s}$ as a function of energy.
  Note that $\B{j}_\text{s}$ changes its sign at higher energies, which explains why manipulating the distribution function can reverse the spin supercurrent~$\B{J}_\text{s}$. 
  {The remaining panels show phase diagrams as functions of the voltage and (c)~tunneling conductance, (d)~polarization, (e)~spin-mixing conductance, and (f)~normal-metal length.
  The remaining parameters are the same as in panel~(a). 
  Yellow regions correspond to significant {antiferro}magnetic interactions ($J_\text{s} > +0.001J_\text{s0}$), and red to significant {ferro}magnetic interactions ($J_\text{s} < -0.001J_\text{s0}$).
  The dashed lines indicate the voltages required to maximize these interactions.}
}
\label{fig:exchange} 
\end{figure}

Recently, the superconducting exchange coupling between ferromagnets was experimentally reported in Ref.~\cite{zhu_natmat_17}.
By lowering the temperature below the superconducting critical temperature $T_\text{c}$, an antiferromagnetic effective exchange interaction was induced by the transition to the superconducting state.
Here, we have shown that the superconducting exchange interaction can be toggled between {anti- and ferro}magnetic via electric voltage, providing a new mechanism compatible with devices operating out-of-equilibrium for actively controlling the magnetic state.
Physically, the sign change of the exchange interaction can be understood from the fact that the voltage alters the occupation of not only states carrying the spectral (energy-resolved) charge supercurrent through the junction, but also the spin supercurrent.

For a more thorough explanation of the effect, we have to consider the spectral spin supercurrents.
The total spin supercurrent~$\B{J}_\text{s}$ can be expressed as an integral 
\begin{equation}
  \B{J}_\text{s} = {J_\text{s0}} \int\limits_0^\infty \d\epsilon\, \B{j}_\text{s}(\epsilon)\,h(\epsilon),
\end{equation}
where the spectral spin supercurrent~$\B{j}_\text{s}$ describes the spin{-}supercurrent-carrying states available, and the distribution function~$h(\epsilon)$ describes which of these are occupied.
According to \cref{eq:dist} in the limit $T \rightarrow 0$, the distribution function at $\epsilon>0$ can be summarized as a step function $\Theta(\epsilon-eV/2)$, where we assume a positive voltage~$V$.
Putting these equations together, we see that the spin supercurrent is basically just an integral of~$\B{j}_\text{s}$ from $\epsilon=eV/2$ and up.
In \cref{fig:exchange}(b), we have plotted the numerically calculated spectral spin supercurrent as a function of energy.
The result is primarily positive for $\epsilon < 0.35\Delta_0$, and primarily negative for $\epsilon > 0.35\Delta_0$.
Since the equation above shows that the voltage~$eV/2$ plays the role of a cutoff that determines which of these {energy regions} contribute to the total spin supercurrent, it becomes clear why the spin supercurrent can be switched via an electric voltage. 
{The mechanism is thus similar to the charge supercurrent switching ~\cite{wilhelm_prl_98} in an S/N/S transistor setup with phase-biased superconductors.}

Our second main result is that the voltage-controlled nonequilibrium quasiparticle distribution can be used to control the anomalous Josephson effect.
{We have for concreteness considered a fixed spin chirality $\chi$ corresponding to perpendicularly oriented magnetization vectors $\B{m}_\textsc{l}, \B{m}_\textsc{r},$ and $\B{m}$. }
\cref{fig:phi0}(a) shows the phase shift as a function of applied voltage. 
As the phase increases from its minimum value $\phi_0 \approx \pi/4$ at $eV/2=0.2\Delta_0$ to a maximum $\phi_0 \approx \pi$ near $eV/2=\Delta_0$, the phase shift is seen to be tuned by more than 120$^\circ$ within a voltage regime of $\sim\! 1$~meV.
It is worth emphasizing that the voltage required here to change the $\phi_0$-shift is two orders of magnitude smaller than the gate voltage $\sim\!200$~meV used in the recent experiment Ref.~\cite{szombati_nphys_15}.
This suggests that the anomalous phase shift proposed in {this} manuscript can be tuned with much less power dissipation than by using gated quantum dots.

The physical mechanism behind the voltage-controlled phase shift can be understood as follows.
The total supercurrent flowing in a Josephson junction with a finite spin-chirality $\chi$ has two contributions according to $J_\text{e} = J_\text{c1}\sin\phi + J_\text{c2}\cos\phi$ where $J_\text{c2} \sim \chi$~\cite{asano_prb_07, grein_prl_09, kulagina_prb_14, eschrig_nphys_08, liu_prb_10, mironov_prb_15, silaev_prb_17}.
The latter term is responsible for the anomalous supercurrent at zero phase difference, as can be seen by rewriting the current-phase relation to the form ${J_\text{e} = J_\text{c}\sin(\phi-\phi_0)}$ where $\phi_0$ depends on the relative magnitude of $J_\text{c1}$ and $J_\text{c2}$.
From previous works considering S/N/S transistors~\cite{wilhelm_prl_98, Volkov1995, baselmans_nature_99}, it is known that the conventional term $J_\text{c1}$ can be forced to change sign by inducing a nonequilibrium energy distribution, corresponding to a 0--$\pi$ transition.
Precisely at this transition point, only the anomalous part $\cos\phi$ remains which is seen in the red curve ($eV/2 = 0.5\Delta_0$) of \cref{fig:phi0}(b).
As one moves away from the 0--$\pi$ transition point, the critical supercurrent may increase since now both $J_\text{c1}$ and $J_\text{c2}$ contribute to $J_\text{e}$.
This matches well with the ${eV/2=0.3\Delta_0}$ and ${eV/2=0.6\Delta_0}$ curves in \cref{fig:phi0}.
Additionally, since the ratio $J_\text{c1}/J_\text{c2}$ changes rapidly around the 0--$\pi$ transition point corresponding to $eV/2=0.5\Delta_0$, we would expect the anomalous phase shift to also vary rapidly near this voltage.
This is confirmed by the results in \cref{fig:phi0}(a).

{We have not included spin-flip and spin-orbit scattering processes in our models.
In general, spin-flip impurities destroys all Cooper pairs, while spin-orbit impurities destroy only triplet pairs.
Thus, at large concentrations, such impurities can be expected to simply reduce the magnitude of both our predicted effects.
How Cooper pairs and supercurrents are affected by these scattering processes is well-known from previous studies, \eg Ref.~\onlinecite{Ouassou2017} and the references therein.}

The electrically tunable anomalous phase shift could be of interest for the purpose of designing a phase battery.
Similarly to how conventional batteries store a potential difference which can drive resistive currents, an anomalous Josephson junction provides a built-in phase difference which could be used to drive supercurrents.
Recent works on magnetic Josephson junctions have taken steps toward realizing such a phase control~\cite{glick_sciadv_18}. 
{{U}nless the magnetic anisotropies of the system are such that a finite spin chirality $\chi$ exists in the ground-state, the misoriented magnetization configuration producing $\chi\neq0$ has to be fixed by external conditions such as an applied magnetic field.}

\begin{figure}[t!]
\centering
\includegraphics[width=\columnwidth]{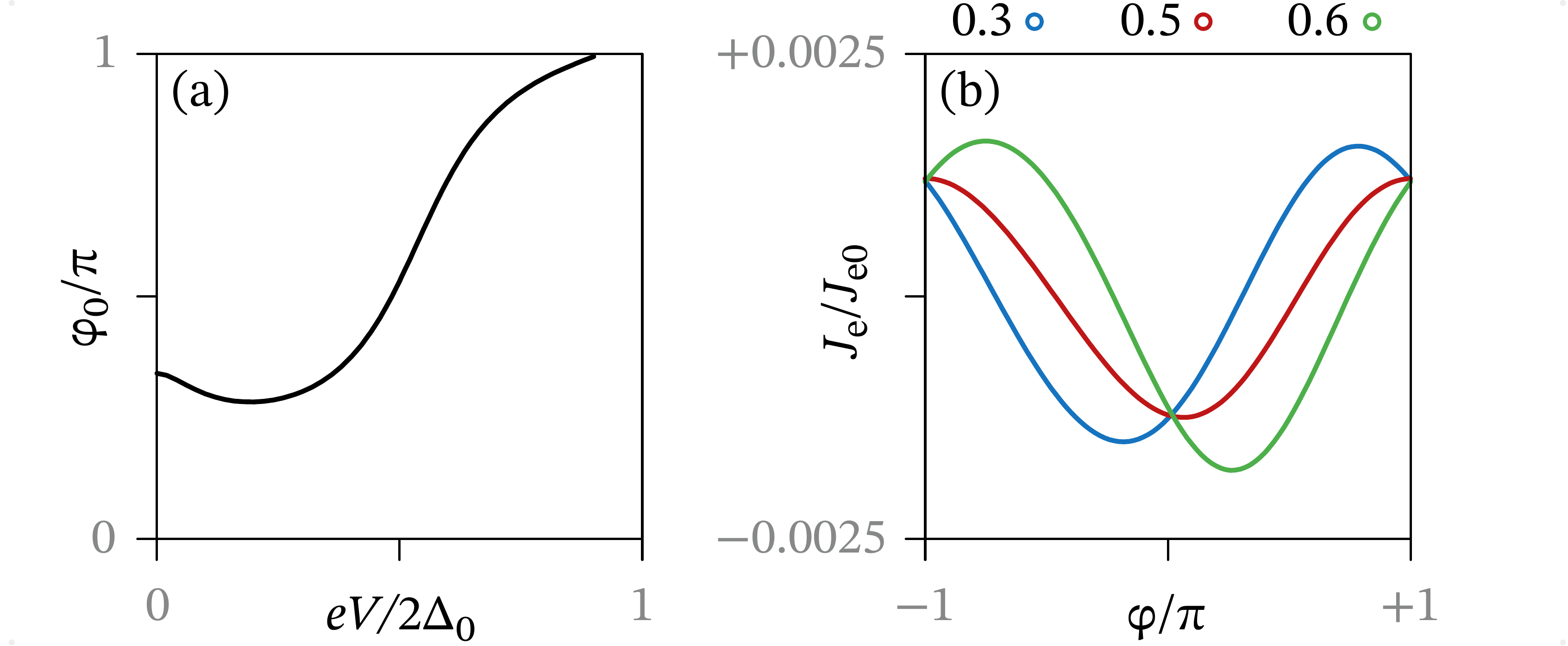}
\caption%
{%
  (Color online) Phase shift causing the anomalous Josephson effect in the system depicted in \cref{fig:model}(b).
  We have used the same interfacial parameter set as in \cref{fig:exchange} and set the length $L_\textsc{f} = 0.5\xi$ and exchange field $m=5\Delta_0$ for the ferromagnetic metal weak link.
}
\label{fig:phi0} 
\end{figure}

\section{Conclusion}
We have investigated the influence of nonequilibrium quasiparticle modes on a superconducting spin-valve and discovered two effects.
First, the voltage-controlled distribution function controls the magnitude and sign of the superconducting exchange interaction, toggling the preferred configuration of the spin-valve from {anti- to ferro}magnetic.
Moreover, we show that the same basic setup controls the anomalous Josephson effect in a junction with finite spin-chirality, obviating the requirement to manipulate the magnetic texture of the system.
We believe this two phenomena may be of interest for the design of nonequilibrium superconducting devices that exploit spin-dependent quantum effects.


\newpage
\begin{acknowledgments}
  We thank O.~Bolme and V.~Risinggård for discussions.
  The numerics was performed on resources provided by UNINETT Sigma2---the national infrastructure for high performance computing and data storage in Norway.
  This work was supported by the Research Council of Norway through grant 240806, and its Centres of Excellence funding scheme grant 262633 ``\emph{QuSpin}''.
\end{acknowledgments}



\begin{thebibliography}{999}

\bibitem{eschrig_rpp_15}
  M. Eschrig.
  Rep. Prog. Phys. \textbf{78}, 104501 (2015).

\bibitem{beckmann_jpcm_16}
  D. Beckmann.
  J. Phys.: Condens. Matter \textbf{28}, 163001 (2016).

\bibitem{bergeret_arxiv_17}
  F.S. Bergeret, M. Silaev, P. Virtanen, and T.T. Heikkilä.
  arXiv:1706.08245.

\bibitem{kalenkov_prl_12}
  M.S. Kalenkov, A.D. Zaikin, and L.S. Kuzmin.
  Phys. Rev. Lett. \textbf{109}, 147004 (2012).

\bibitem{machon_prl_13}
  P. Machon, M. Eschrig, and W. Belzig.
  Phys. Rev. Lett. \textbf{110}, 047002 (2013).

\bibitem{kolenda_prl_16}
  S. Kolenda, M.J. Wolf, and D. Beckmann.
  Phys. Rev. Lett. \textbf{116}, 097001 (2016).

\bibitem{wakamura_natmat_15}
  T. Wakamura, H. Akaike, Y. Omori, Y. Niimi,  S. Takahashi, A. Fujimaki, S. Maekawa, and Y. Otani.
  Nature Materials \textbf{14}, 675–678 (2015).

\bibitem{ouassou_arxiv_18}
  J.A. Ouassou, T.D. Vethaak, and J. Linder.
  Phys. Rev. B \textbf{98}, 144509 (2018).
	
	\bibitem{bobkova_prb_11} I. V. Bobkova and A. M. Bobkov.
	Phys. Rev. B. \textbf{84}, 140508(R) (2011).

\bibitem{waintal_prb_02}
  X. Waintal and P.W. Brouwer.
  Phys. Rev. B \textbf{65}, 054407 (2002).

\bibitem{DeGennes1966}
  {P.G. de\,Gennes. Physics Letters \textbf{23}, 10 (1966)}

\bibitem{Manske2014}
  {W.~Chen, P.~Horsch, and D.~Manske. Phys. Rev. B \textbf{89}, 064427 (2014).}

\bibitem{zhao_prb_08}
  E. Zhao and J.A. Sauls.
  Phys. Rev. B \textbf{78}, 174511 (2008).

\bibitem{linder_prb_11}
  J. Linder and T. Yokoyama.
  Phys. Rev. B \textbf{83}, 012501 (2011).

\bibitem{holmqvist_prb_12}
  C. Holmqvist, W. Belzig, and M. Fogelström.
  Phys. Rev. B \textbf{86}, 054519 (2012).

\bibitem{takashima_prb_17}
  R. Takashima, S. Fujimoto, and T. Yokoyama.
  Phys. Rev. B \textbf{96}, 121203(R) (2017).

\bibitem{bobkova_prb_18}
  I.V. Bobkova, A.M. Bobkov, and M.A. Silaev.
  Phys. Rev. B \textbf{98}, 014521 (2018).

\bibitem{linder_nphys_15}
  J. Linder and J.W.A. Robinson.
  Nature Physics \textbf{11}, 307 (2015).

\bibitem{baselmans_nature_99}
  J.J.A. Baselmans, A.F. Morpurgo, B.J. van Wees, and T.M. Klapwijk.
  Nature \textbf{397}, 43 (1999).

\bibitem{wilhelm_prl_98}
  F.K. Wilhelm, G. Schön, and A.D. Zaikin.
  Phys. Rev. Lett. \textbf{81}, 1682 (1998).

\bibitem{Volkov1995}
  A.F.~Volkov.
  Phys. Rev. Lett. 74, 4730 (1995).
	
\bibitem{slon_prb_89}
  J.C. Slonczewski.
  Phys. Rev. B \textbf{39}, 6995 (1989).

\bibitem{nogueira_epl_04}
  F.S. Nogueira and K.-H. Bennemann. 
  EPL \textbf{67}, 620 (2004).

\bibitem{szombati_nphys_15}
  D.B. Szombati, S. Nadj-Perge, D. Car, S.R. Plissard, E.P.A.M. Bakkers, L.P. Kouwenhoven. 
  Nat. Phys. \textbf{12}, 568 (2016).

\bibitem{glick_sciadv_18}
  J.A. Glick \etal.
  Sci. Adv. \textbf{27}, 7 (2018).

\bibitem{heikkila_epl_00}
  T.T. Heikkilä, W.K. Wilhelm, and G. Schön.
  EPL \textbf{51}, 434 (2000).

\bibitem{Chandrasekhar2008}%
  V.~Chandrasekhar.
  In \emph{Superconductivity} (Springer, Berlin, Heidelberg, 2008)\ pp.\ 279.

\bibitem{Belzig1999}%
  W.~Belzig, F.K.~Wilhelm, C.~Bruder, G.~Schön, and A.D. Zaikin.
  Superlattices and Microstructures \textbf{25}, 1251 (1999).

\bibitem{Rammer1986}%
  J.~Rammer and H.~Smith.
  Rev. Mod. Phys. \textbf{58}, 323 (1986).

\bibitem{Spinject}
  {J.A.~Ouassou, J.W.A.~Robinson, J.~Linder.
  arXiv:1810.08623.
  }

\bibitem{Usadel1970}%
  K.D.~Usadel.
  Phys. Rev. Lett. \textbf{25}, 507 (1970).

\bibitem{kontos_prl_01}
  T. Kontos, M. Aprili, J. Lesueur, and X. Grison.
  Phys. Rev. Lett. \textbf{86}, 304 (2001).

\bibitem{Schopohl}%
  N.~Schopohl.
  arXiv:cond-mat/9804064.

\bibitem{SpinOrbit}
  S.H.~Jacobsen, J.A.~Ouassou, and J.~Linder.
  Phys. Rev. B \textbf{92}, 024510 (2015).

\bibitem{Halfmetal}%
  {
    J.A.~Ouassou, A.~Pal, M.~Blamire, M.~Eschrig, J.~Linder.
    Sci. Rep. \textbf{7}, 1932 (2017).
  }

\bibitem{Eschrig2015b}%
  M.~Eschrig, A.~Cottet, W.~Belzig, and J.~Linder.
  New J. Phys. \textbf{17}, 083037 (2015).

\bibitem{Cottet2009}%
  A.~Cottet, D.~Huertas-Hernando, W.~Belzig, and Y.V. Nazarov.
  Phys. Rev. B \textbf{80}, 184511 (2009).

\bibitem{Cottet2007a}%
  A.~Cottet. 
  Phys. Rev. B \textbf{76}, 224505 (2007).

\bibitem{Bobkova2015a}%
  I.V.~Bobkova and  A.M. Bobkov.
  JETP Lett. \textbf{101}, 118 (2015).

\bibitem{Silaev2015}%
  M.~Silaev, P.~Virtanen, F.S.~Bergeret, and T.T.~Heikkilä.
  Phys. Rev. Lett. \textbf{114}, 167002 (2015).

\bibitem{Aikebaier}%
  F.~Aikebaier, M.A.~Silaev, and T.T.~Heikkilä.
  arXiv:1712.08653.

\bibitem{Pothier1997}%
  H.~Pothier, S.~Guéron, N.O.~Birge, D.~Esteve, and M.H.~Devoret.
  Phys. Rev. Lett. \textbf{79}, 3490 (1997).

\bibitem{zhu_natmat_17}
  Y. Zhu, A. Pal, M. Blamire, and Z.H. Barber.
  Nature Materials \textbf{16}, 195 (2017).

\bibitem{kulagina_prb_14}
  I. Kulagina and J. Linder.
  Phys. Rev. B \textbf{90}, 054504 (2014).

\bibitem{asano_prb_07}
  Y.~Asano, Y.~Sawa, Y.~Tanaka, and A.~Golubov.
  Phys. Rev. B \textbf{76}, 224525 (2007).

\bibitem{grein_prl_09}
  R.~Grein, M.~Eschrig, G.~Metalidis, and G.~Schön.
  Phys. Rev. Lett. \textbf{102}, 227005 (2009).
	
\bibitem{eschrig_nphys_08} 
  M.~Eschrig and T.~Löfwander.
  Nat. Phys. \textbf{4}, 138 (2008).
	
\bibitem{liu_prb_10} 
  J.-F.~Liu and K.S.~Chan.
	Phys. Rev. B \textbf{82}, 184533 (2010).
	
\bibitem{mironov_prb_15} 
  S.~Mironov and A.~Buzdin.
	Phys. Rev. B \textbf{92}, 184506 (2015).
	
\bibitem{silaev_prb_17} 
  M.A.~Silaev, I.V.~Tokatly, and F.S.~Bergeret. 
  Phys. Rev. B \textbf{95}, 184508 (2017).

\bibitem{Ouassou2017}
  {J.A.~Ouassou, S.H.~Jacobsen, and J.~Linder.
  Phys. Rev. B \textbf{96}, 094505 (2017).}

\end{thebibliography}
\end{document}